\documentstyle[preprint,prb,aps]{revtex}

\tighten

\begin{document}

\draft

\preprint{MA/UC3M/12/1994}

\title{Quasi-ballistic electron transport in random superlattices}

\author{Francisco Dom\'{\i}nguez-Adame}

\address{Departamento de F\'{\i}sica de Materiales,
Facultad de F\'{\i}sicas, Universidad Complutense,
E-28040 Madrid, Spain}

\author{Angel S\'anchez and Enrique Diez}

\address{Escuela Polit\'ecnica Superior,
Universidad Carlos III de Madrid,
C./ Butarque, 15, E-28911 Legan\'es, Madrid, Spain}

\maketitle

\begin{abstract}

We theoretically study electron transport in disordered, quantum-well
based, semiconductor superlattices with structural short-range
correlations.  Our system consists of equal width square barriers and
quantum wells with two different thicknesses.  The two kinds of quantum
wells are randomly distributed along the growth direction.  Structural
correlations are introduced by adding the constraint that one of the
wells always appears in pairs.  We show that such correlated disordered
superlattices exhibit a strong enhancement of their dc conductance as
compared to usual random ones, giving rise to quasi-ballistic electron
transport.  Our predictions can be used to demonstrate experimentally
that structural correlations inhibit the localization effects of
disorder.  We specifically describe the way superlattices should be
built and experiments should be carried out for that purpose.

\end{abstract}


\narrowtext

\section{Introduction}

It is by now well established on firm theoretical grounds that electron
localization may be suppressed and bands of extended states appear in
one-dimensional random systems with structural short-range correlations
(see, e.g., Refs.\ \onlinecite{Wu3} and \onlinecite{JPA} and references
therein).  This unexpected phenomenon is not restricted to electronic
systems but rather seems to be quite general, as it is has also dramatic
effects in classical harmonic chains \cite{nozotro,indios2}, magnon
propagation, \cite{Evan2} or exciton dynamics.\cite{exciton} All these
theoretical analyses contradict the earlier belief that {\em all}
eigenstates are localized in one-dimensional disordered
systems.\cite{Ziman} Due to the lack of experimental confirmation, there
is still some controversy as to the relevance of these results and their
physical implications on transport properties.  Here we concern
ourselves with semiconductor superlattices (SLs) in order to find {\em
experimentally measurable quantities and physically realizable systems}
that allow for a clearcut validation of the above mentioned results.
The reason for the choice of SLs for this purpose is twofold: First,
they have been already used succesfully to observe electron localization
in intentionally uncorrelated disordered quantum-well (QW) based
GaAs/Ga$_{1-x}$Al$_x$As SLs. \cite{Chomette} On the other hand, previous
results of us \cite{Diez} on simple, highly idealized models of SLs
indicate that the effects of correlated disordered should be clearly
visible in such systems.  These two reasons, and the fact that a number
of studies have been performed to date on carrier dynamics in disordered
SLs, \cite{Sasaki} make these systems the ideal candidates to propose
experiments on localization or delocalization electronic properties.
Accordingly, in this Rapid Communication we deal with
GaAs/Ga$_{1-x}$Al$_x$As SLs specifically designed for experiments on
correlated disorder.  In Sec.~II we present our system and our
analytical results on transport properties.  Section~III contains our
numerical studies on electron transport through disordered SLs and the
effect we will call {\em quasi-ballistic electron transport\/}.
Finally, in Sec.~IV we discuss our results.

\section{Analytical results}

\subsection{Electron dynamics in QWSL\lowercase{s}}

The SL consists of two kinds of semiconductors layers (hereafter called
$A$ and $B$) arranged alternatively along the growth direction $X$.  Let
$\Delta E_c$ be the conduction-band offset defined as $E_{cB}-E_{cA}$
and without loss of generality we will take $\Delta E_c>0$.  In
addition, we further consider that the thickness of layers $B$ is the
same in the whole SL and denote it by $b$.  Denoting by $x_n$ the
spatial coordinate of the center of the {\em n\/}th barrier, the
conduction-band profile is given as $V_{SL}(x)=\sum_n V(x-x_n)$, where
$V(x-x_n)$ takes the constant value $\Delta E_c$ for $|x-x_n|<b/2$ and
vanishes otherwise.  We focus on electron states close to the band gap
with ${\bf k}_{||}=0$ so we can use the one-band effective-mass
framework to calculate the envelope-functions
\begin{equation}
\label{Schr}
\left[-\,{\hbar^2\over 2m^*}\,{d^2\phantom{x}\over dx^2} +
V_{SL}(x) \right] \psi(x) = E\>\psi(x),
\end{equation}

where the explicit dependence of both $E$ and $\psi$ on quantum numbers
is understood and they will be omited in the rest of the paper.  We have
taken a constant effective-mass $m^*$ at the $\Gamma$ valley although
this is not a serious limitation as our description can be easily
generalized to include two different efective masses.  Let us consider
states below the barrier, which are of most interest to study quantum
confinement effects.  The corresponding envelope-function values at both
sides of a barrier are related via a $2\times 2$ transfer-matrix $M(n)$
whose elements are $M_{11}(n)=M_{22}^*(n)\equiv \alpha_n$ and
$M_{12}(n)=M_{21}^*(n)\equiv \beta_n$ where
\begin{mathletters}
\label{elements}
\begin{eqnarray}
\alpha_n & = & \left[ \cosh(\eta b) +i\, \left( {\kappa^2-\eta^2
\over 2\kappa\eta} \right) \sinh (\eta b) \right]
\exp [ i\kappa (\Delta x_n -b) ] \label{elementsa} \\
\beta_n & = & -i\, \left( {\kappa^2+\eta^2
\over 2\kappa\eta} \right) \sinh (\eta b)\,
\exp [ -i\kappa (\Delta x_n -b) ] \label{elementsb},
\end{eqnarray}
\end{mathletters}
with $\Delta x_n \equiv x_n-x_{n-1}$, $\kappa^2=2m^*E/\hbar^2$ and
$\eta^2 = 2m^*(V-E)/\hbar^2$.  Letting $N$ be the total number of
barriers, the transfer matrix of the SL is obtained as the product
$T(N)=M(N)MN(N-1)\cdots M(1)$.  The element $T_{11}(N)\equiv A_N$ can be
easily calculated recursively from the relationship \cite{JPA}
\begin{equation}
A_n=\left( \alpha_n + \alpha_{n-1}^*{\beta_n \over \beta_{n-1}} \right)
A_{n-1} -\,\left( {\beta_n \over \beta_{n-1}} \right) A_{n-2},
\label{A}
\end{equation}
supplemented by the initial conditions $A_0=1$, $A_1=\alpha_1$.  The
knowledge of $A_N$ enables us to obtain the transmission coefficient
$\tau$ at a given energy $E$, $\tau=|A_N|^{-2}$, and the single-channel,
dimensionless Landauer resistance, $\rho = 1/\tau-1=|A_N|^2-1$.
Finally, the dimensionless Lyapunov coefficient is a nonnegative
parameter given by \cite{Kirk} $\gamma=-\,{1\over 2N} \ln \tau$, being
nothing but the inverse of the localization length.

\subsection{Transmission through a single DQW}

We now consider a single dimer quantum well (DQW), with the {\em k\/}th
barrier in between, in an otherwise periodic SL. We denote the thickness
of the QW in the periodic SL by $a$ whereas the thickness of each QW
forming the DQW is denoted by $a'$, as shown in Fig.~\ref{fig0}.  The
condition for an electron to move in the periodic SL is
$|\mbox{Tr}[M(1)]|\leq 2$ and the correspoding minibands are
\begin{equation}
\left| \cos (\kappa a) \cosh (\eta b)-\left( {\kappa^2-\eta^2\over
2\kappa\eta}\right) \sin (\kappa a) \sinh (\eta b) \right| \leq 1.
\label{condicion1}
\end{equation}
For brevity we put $\alpha_k=\alpha_{k+1}\equiv \alpha'$ and
$\alpha_n \equiv \alpha$ ($n\neq k, k+1$). Considering Eq.~(\ref{A})
for $n=k, k+1, k+2$, eliminating $A_k$ and $A_{k+1}$ and setting
$\mbox{\rm Re}(\alpha')=0$ we obtain after a little algebra
\begin{equation}
-A_{k+2}=(\alpha+\alpha^*)A_{k-1} -A_{k-2}.
\label{AA}
\end{equation}
Besides a constant phase factor of $\pi$ which has no effects on the
magnitudes of interest, Eq.~(\ref{AA}) reduces to Eq.~(\ref{A}) for a
periodic SL in which sites $k$ and $k+1$ has been eliminated.  This
means that the reflection coefficient at the DQW vanishes and,
consequently, there exists complete transparency at the resonant energy
$E_r$ satisfying $\mbox{\rm Re}(\alpha')=0$, i.\ e.,
\begin{equation}
\cos (\kappa_r a') \cosh (\eta_r b)-\left( {\kappa_r^2-\eta_r^2\over
2\kappa_r\eta_r}\right) \sin (\kappa_r a') \sinh (\eta_r b)=0,
\label{condicion2}
\end{equation}
where the subscript $r$ refers to the resonant energy $E_r$.
Importantly, choosing $a'$ appropriately allows to locate the resonant
energy $E_r$ within an allowed miniband of the periodic SL, that is,
the resonant energy in the range of energies given by
Eq.~(\ref{condicion1}).

\section{Transport through a DQWSL}

We now turn to the problem we are interested in, namely SLs with a
finite concentration of DQWs, to verify whether the single DQW resonance
is still preserved.  To this end, we apply the previous results to a
specific case, namely GaAs/Ga$_{0.65}$Al$_{0.35}$As.  In this case
$\Delta E_c =0.25\,$eV and $m^*=0.067m$, $m$ being the electron mass.
In our computations we have taken $a=b=200\,$\AA\ and $a'=160\,$\AA.
With these parameters we find from Eq.~(\ref{condicion1}) only one
allowed miniband below the barrier, ranging from $0.1022\,$eV up to
$0.1755\,$eV.  The resonant energy is $E_r=0.1565\,$eV from
Eq.~(\ref{condicion2}) and thus it lies in this allowed miniband.  The
maximum number of barriers we have considered is $N=1000$ and the number
of wells with thickness $a'$ is $N/5$, althogh we have cheked that the
main conclusions of the present work are independent of this ratio.  We
have generated random SLs with and without the constraint of pairing,
but always with the same number of wells of thickness $a'$.  The
physical magnitudes we are interested in were averaged for several
realizations of the SLs. The ensembles comprised a number of
realizations varying from $200$ up to $400$ to test the convergence of
the computed mean values, and this convergence was always satisfactory.

The transmission coefficient around the resonant energy is shown in
Fig.~\ref{fig1} for SLs with $N=500$ barriers.  We stress that, in spite
of the fact that the plot corresponds to an average over $300$
realizations of the SLs, the transmission coefficient for typical
realizations behaves in the same way, although noisier.  Close to the
resonant energy there is an interval of energies that shows also very
good transmission properties, similar to those of the resonant energy.
This strong peak is not observed when DQW are absent.  Such peak implies
the appearance of a deep minimum in the Landauer resistance close to
$E_r$, as it becomes evident from the relationship between $\tau$ and
$\rho$.  For brevity we do not show the corresponding figure, but it is
worth mentioning that there is several order of magnitudes between the
values of the resistance close to $E_r$ when DQW are present or not.

In Fig.~\ref{fig2} we present the size dependence results for three
different energy values (all of them lie in the allowed miniband) in
random SLs with DQW. For those states with energy at the resonant one
$E_r$ the behavior is perfectly omhic, presenting only small
fluctuations around the mean value.  On the contrary, when we separate
from the resonant energy we observe a nonhomic behavior of the
resistance increasing exponentially with the system size: the more
distant from the resonant energy, the larger the exponential growth of
the resistance with the system size.  The fact that around the resonant
energy there is good transmission, also revealing itself in an omhic
behavior of the Landauer resistance in spite of the randomness, suggests
the possibility that the localization length of those states may be very
large.  This is shown through the Lyapunov exponent in Fig.~\ref{fig3}
for the same SLs as in Fig.~\ref{fig1}.  The comparison between random
SLs with and without DWQ is actually dramatic in the whole range of
energies shown in Fig.~\ref{fig3}, and reflects the property that a
large number of states close to the resonant energy have a very large
localization length.  In fact those states spread over the whole
superlattice, thus justifying to refer them as to {\em quasi-ballistic}
states.

\section{Conclusions}

We have studied electron transport in QW-based random SLs with and
without DQW, showing that there exists a resonant energy for which a
complete transparency through a single DWQ is achieved.  This resonant
energy depends only on structural parameters (layer thicknesses) in a
given SL and, consequently, it is possible to place it within a miniband
of the periodic SL. As a major point, we have found that these resonance
effects also arise when a finite number of DQW are randomly placed in
the SL, leading to very good transmission in a finite energy range
around the resonant one.  In a simple Kronig-Penney model with dimer
impurities we have previously found \cite{Diez} that such minimum cause
a dramatic enhacement of the dc conductivity at finite temperature
whenever the Fermi level lies close to it.  Our present results strongly
support the idea that similar effects should be experimentaly observable
in actual SLs with correlated disorder.  Moreover, states close to the
resonant energy present omhic behavior, whereas when we deviate from
this energy the resistance shows an exponential increase with the system
size.  Finally, such {\em quasi-ballistic} electron states present a
very large localization length, opposite to what occurs in random SLs
without the constraint of pairing.

Experiments on such SLs would validate (or discard) all the recent
claims that correlation induces the appearance of extended states in
spite of disorder.  A possible experimental setup is as follows.  The
random SL is inserted between two thick barriers doped with a high
density of Si (tipically $1\times 10^{18}\,$cm$^{-3}$), so that the
Fermi level is pinned at the dopant energy level (about $0.14\,$eV in
Ga$_{0.65}$Al$_{0.35}$As).  Different random SLs are prepared varying
$a'$ while keeping constant $a$ and $b$ (say $a=b=200\,$\AA\ as those
SLs we have studied).  By varying the value of $a'$ the resonant energy
is moved through the allowed miniband (for instance, $E_r$ ranges from
$0.1187\,$eV for $a'=220\,$\AA\ up to $0.1643\,$eV for $a'=150\,$\AA.).
Therefore, plots of the SL dc conductance at low temperature as a
function of $a'$ should exhibit a clear peak when the resonant energy
matches the Fermi level.  If this maximum is actually observed we will
then be led to the conclusion than quasi-ballistic transport is taking
place.

We hope that our results may encourage experimental effort in this
direction for three reasons.  First, and most importantly, to validate
or not the existence of extended states in random systems with
short-range correlations.  Second, to elucidate whether the
molecular-beam-epitaxy techniques can fabricate samples with the
required perfection such that quantum coherence is not destroyed and
resonance effects may arise.  Finally, the feature of having
quasi-ballistic transport only for certain energies may be the basis for
designing new devices and applications.

\acknowledgments

A.\ S.\ is thankful to Alan Bishop for warm hospitality at Los Alamos
National Laboratory where this paper was written in part.  Work at
Madrid is supported by UCM through project PR161/93-4811.  Work at
Legan\'es is supported by the DGICyY (Spain) through project PB92-0248,
and by the European Union Human Capital and Mobility Programme through
contract ERBCHRXCT930413.

\begin{figure}
\caption{Schematic diagram of the SL with a single DQW. Shaded area
represents an allowed miniband in the periodic SL and the thick line
stands for the resonant level in the DQW. Labels indicate the thickness
of layers.}
\label{fig0}
\end{figure}

\begin{figure}
\caption{Transmission coefficient around the resonant energy for a
random QWSL (upper curve) and a DQWSL (lower curve).  Shown are averages
over 300 realizations, every SL consists of $N=400$ barriers of
$b=200\,$\AA\ whereas the thicknesses of QW are $a=200\,$\AA\ and
$a'=160\,$\AA.}
\label{fig1}
\end{figure}

\begin{figure}
\caption{Landauer resistance as a function of the number of barriers in
GaAs/Ga$_{0.65}$Al$_{0.35}$As SLs with DQW for different energies: the
resonant one $E_r=0.1565\,$eV (lower curve), $0.9E_r=0.1409\,$eV (middle
curve), and $0.8E_r=0.1252\,$eV (upper curve).  Parameters are the same
as in Fig.~\protect{\ref{fig1}}.}
\label{fig2}
\end{figure}

\begin{figure}
\caption{Lyapunov coefficient around the resonant energy for a random
QWSL (upper curve) and a DQWSL (lower curve).  Parameters are the same
as in Fig.~\protect{\ref{fig1}}.}
\label{fig3}
\end{figure}

\end{document}